\documentclass{ckm}                 

\confname{Workshop on the CKM Unitarity Triangle, IPPP Durham, April
  2003}

\title{Evidence for $B^+\rightarrow\omega l^+\nu$}

\author{C.\ Schwanda\thanks{Supported by the Japan Society for the
    Promotion of Science.} (for the Belle collaboration)}
\address{High Energy Accelerator Research Organization (KEK), Tsukuba,
    Japan}

\begin{document}

\begin{abstract}
  We have searched for the decay~$B^+\rightarrow\omega l^+\nu$ in
78~fb$^{-1}$ of $\Upsilon(4S)$ data (85.0~million $B\bar B$ events)
accumulated with the Belle detector. The final state is fully
reconstructed using the $\omega$~decay into $\pi^+\pi^-\pi^0$ and the
detector hermeticity to infer the neutrino momentum. The signal yield
is extracted by a two-dimensional fit to the lepton momentum and the
invariant $\pi^+\pi^-\pi^0$~mass. The result of the fit depends on the
form factor model assumed for the decay. Taking the average over three
different models, $421\pm 132$~events are found in the data,
corresponding to a preliminary branching fraction of $(1.4\pm
0.4(stat)\pm 0.2(syst)\pm 0.3(model))\cdot 10^{-4}$.

\end{abstract}

\maketitle

\setcounter{footnote}{0}

\section{Introduction}

Tree-level semileptonic $B$~decays are crucial for
measuring the CKM matrix elements $|V_{cb}|$ and $|V_{ub}|$~\cite{ref:1}
independently of any new physics contribution. The
decay~$B\rightarrow X_ul\nu$~\footnote{Throughout this article, the
  inclusion of the charge conjugate mode is implied.} (which
allows to access $|V_{ub}|$) has been studied both using inclusive
approaches (sensitive to all $X_ul\nu$~final states within a given
region of phase space) and through exclusive reconstruction of
specific final states~\cite{ref:2}. As to the latter, Belle has
already obtained preliminary results for the decays
$B^0\rightarrow\pi^-l^+\nu$ and
$B^+\rightarrow\rho^0l^+\nu$~\cite{ref:3}. In this article, we present
evidence for the decay~$B^+\rightarrow\omega l^+\nu$ which is so far
unobserved.

\section{Experimental procedure}

\subsection{KEKB and the Belle detector}

Belle is located at the KEKB~asymmetric $e^+e^-$~collider, operating
at the $\Upsilon(4S)$~resonance~\cite{ref:4}. The detector is
described in detail elsewhere~\cite{ref:5}. It is a large-solid-angle
magnetic spectrometer that consists of a three-layer silicon vertex
detector (SVD), a 50-layer central drift chamber (CDC), an array of
aerogel threshold \v{C}erenkov counters (ACC), a barrel-like
arrangement of time-of-flight scintillation counters (TOF), and an
electromagnetic calorimeter comprised of CsI(Tl) crystals (ECL)
located inside a super-conducting solenoid coil that provides a 1.5~T
magnetic field.

The responses of the ECL, CDC ($dE/dx$) and ACC detectors are combined
to provide clean electron identification. Muons are identified in
the instrumented iron flux-return (KLM) located outside of the
coil. Charged hadron identification relies on information from
the CDC, ACC and TOF.

\subsection{The data set}

This analysis is based on an integrated luminosity of 78.13~fb$^{-1}$
taken on the $\Upsilon(4S)$~resonance. This data set contains
$(85.0\pm 0.5)\cdot 10^{6}$~$B\bar B$~events. Furthermore, to subtract
the hadronic background, 8.83~fb$^{-1}$ taken below the resonance are
used.

Full detector simulation is applied to Monte Carlo events. Generic
background Monte Carlo samples, equivalent to about three
times the beam luminosity, are used in various stages of this
analysis. Monte Carlo samples for the signal, $B^+\rightarrow\omega
l^+\nu$, are generated with the following form factor models: ISGW2
(quark model~\cite{ref:6}), UKQCD (quenched lattice QCD
calculation~\cite{ref:7}) and LCSR (light cone sum
rules~\cite{ref:8}). The cross-feed from other decays~$B\rightarrow
X_ul\nu$ is estimated using the ISGW2 Monte Carlo.

\subsection{Neutrino reconstruction}

Events passing the hadronic selection are required to contain a single
lepton (electron or muon) with a c.m.\ momentum~$p^*_l$ greater than
1.3~GeV/$c$. In this momentum range, electrons (muons) are selected
with an efficiency of 92\% (89\%) and a pion fake rate of 0.25\%
(1.4\%). This requirement is 41\% efficient for $B\rightarrow
X_ul\nu$~events.

The missing four-momentum is computed for selected events,
\begin{eqnarray}
  \vec p_{miss} & = & \vec p_{HER}+\vec p_{LER}-\sum_i\vec p_i~,
  \nonumber \\
  E_{miss} & = & E_{HER}+E_{LER}-\sum_i E_i~,
\end{eqnarray}
where the sum runs over all reconstructed charged tracks and
photons. The indices HER and LER refer to the high energy and the low
energy ring, respectively. To reject events in which the
missing momentum does not match the neutrino momentum, the following
selection criteria are applied. Events with a large charge imbalance
are eliminated, $|Q_{tot}|<3e$, and the direction of the missing
momentum is required to lie within the ECL~acceptance,
$17^\circ<\theta_{miss}<150^\circ$. The missing mass squared,
$m^2_{miss}=E^2_{miss}-\vec p^2_{miss}$, is required to be consistent
with the neutrino hypothesis, $|m^2_{miss}|<3$~GeV$^2$/$c^4$.

\begin{figure}
  \begin{center}
    \includegraphics{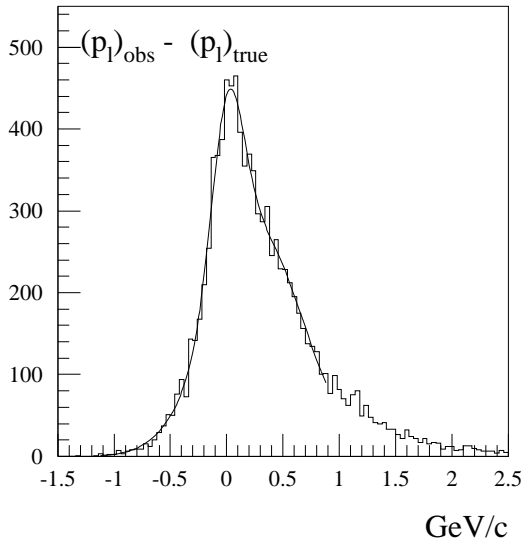}
  \end{center}
  \caption{The missing momentum resolution in $B\rightarrow
    X_ul\nu$~events (ISGW2~model). The fit shown uses a double
    Gaussian ($\sigma_1=141\pm 10$~MeV/$c$, $\sigma_2=412\pm
    8$~MeV/$c$).} \label{fig:1}
\end{figure}

After applying these cuts, the resolution in $p_{miss}$ ($E_{miss}$)
is found to be 141~MeV/$c$ (587~MeV) for simulated $B\rightarrow
X_ul\nu$~events (Fig.~\ref{fig:1}). As the energy resolution is worse
than the momentum resolution, the neutrino four-momentum is taken to
be $(p_{miss},\vec p_{miss})$. The efficiency of the event selection
and neutrino reconstruction cuts is about 17\% for $B\rightarrow
X_ul\nu$~events.

\subsection{Final state reconstruction}

Pairs of $\gamma$'s are combined to form
$\pi^0$~candidates ($E_\gamma>30$~MeV,
$120<m(\gamma\gamma)<150$~MeV/$c^2$). The
decay~$\omega\rightarrow\pi^+\pi^-\pi^0$ (branching ratio: $(89.1\pm
0.7)$\%~\cite{ref:9}) is reconstructed by trying all possible
combinations of one $\pi^0$ with two oppositely charged
tracks. Combinations with a charged track identified as a kaon are
rejected and the following selections are imposed:
$p^*_\omega>300$~MeV/$c$,
$703<m(\pi^+\pi^-\pi^0)<863$~MeV/$c^2$. Combinations located far away
from the center of the Dalitz plot are removed by requiring the Dalitz
amplitude, $A\propto |\vec p_{\pi^+}\times \vec p_{\pi^-}|$, to be
larger than half of its maximum value.

The single lepton in the event is combined with the $\omega$~candidate
and the inclusive neutrino, and the lepton momentum cut is tightened,
$1.5<p^*_l<2.7$~GeV/$c$. To reject combinations inconsistent with
signal decay kinematics, the cut
$|\cos\theta_{BY}|<1.1$ is imposed,
\begin{equation}
  \cos\theta_{BY}=\frac{2E^*_B E^*_Y-m^2_B-m^2_Y}{2p^*_Bp^*_Y}~,
\end{equation}
where $E^*_B$, $p^*_B$ and $m_B$ are fixed to their nominal values,
$\sqrt{E_{HER}E_{LER}}$, $\sqrt{E^{*2}_B-m^2_B}$ and 5.279~GeV/$c^2$,
respectively. $E^*_Y$, $p^*_Y$ and $m_Y$ are the measured c.m.\
energy, momentum and mass of the $Y=\omega+l$~system,
respectively. For well-reconstructed signal events, $\cos\theta_{BY}$
is the cosine of the angle between the $B$ and the $Y$~system
and lies between $-1$ and $+1$ while for the various backgrounds, a
significant fraction is outside this interval.

For each $\omega l\nu$~candidate, the beam-constrained mass $m_{bc}$
and $\Delta E$ are calculated ($E^*_{beam}=\sqrt{E_{HER}E_{LER}}$),
\begin{eqnarray}
  m_{bc} & = & \sqrt{(E^*_{beam})^2-|\vec p^*_\omega+\vec p^*_l+\vec
    p^*_\nu |^2}~, \nonumber \\
  \Delta E & = & E^*_{beam}-(E^*_\omega+E^*_l+E^*_\nu)~,
\end{eqnarray}
and candidates in the window $m_{bc}>5.23$~GeV/$c^2$ and
$|\Delta E|<0.36$~GeV are selected. On the average, two combinations
per event remain after applying all cuts. We choose the one with the
largest $\omega$~momentum in the c.m.\ frame, which is the right
choice about 80\% of the time.

\subsection{Continuum suppression}

The hadronic continuum from $e^+e^-\rightarrow q\bar q$~events,
$q=u,d,s,c$, is suppressed using three variables exploiting the fact
that, in the $\Upsilon(4S)$ frame, the two $B$~mesons are produced
nearly at rest and that therefore $B\bar B$~events have a nearly
spherical shape while continuum events have a more jet-like
topology. These variables are:
\begin{itemize}
  \item The ratio~$R_2$ of the second to the zeroth Fox-Wolfram
    moment~\cite{ref:10}. This ratio tends to be close to zero (unity)
    for spherical (jet-like) events. (The cut $R_2<0.4$ is imposed
    already at the event selection level.)
  \item The cosine of $\theta_{thrust}$, where $\theta_{thrust}$ is
    the angle between the thrust axis of the $\omega l$~system and the
    thrust axis of the rest of the event.
  \item A Fisher discriminant selecting events with an even energy
    distribution around the lepton direction~\cite{ref:11}. The input
    variables are the charged and neutral energy in nine cones of
    equal solid angle around the lepton momentum axis.
\end{itemize}

To optimize their individual discriminating power, these three
variables are combined into a likelihood ratio. This
selection is 55\%~efficient for $B^+\rightarrow\omega l^+\nu$ events,
while 92\% of the continuum background remaining after the
$R_2<0.4$~cut, is eliminated.

\subsection{The fit}

\begin{figure*}
  \begin{center}
    \includegraphics{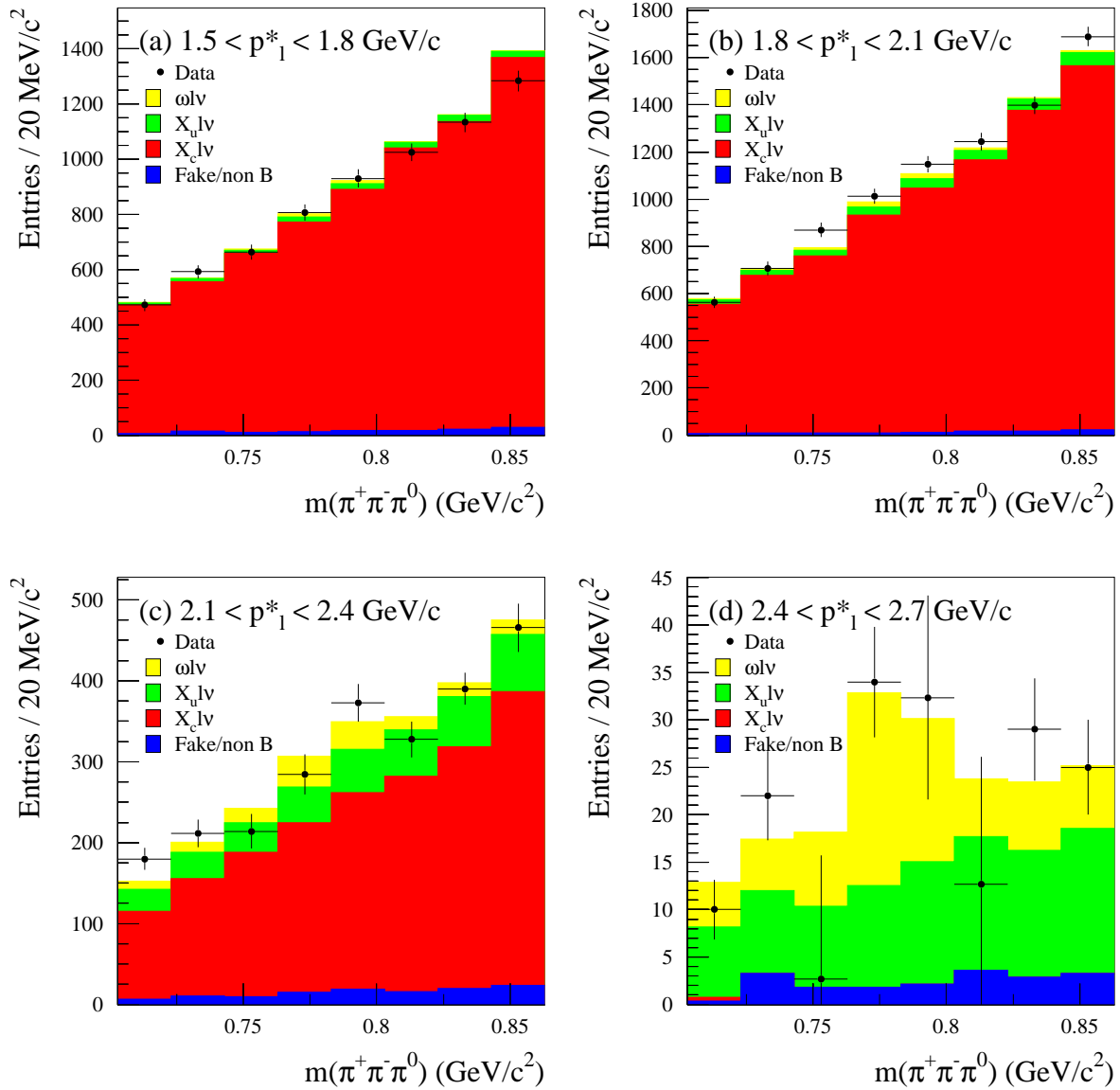}
  \end{center}
  \caption{The result of the fit in bins of $p^*_l$, using ISGW2~form
      factors for the decay $B^+\rightarrow\omega l^+\nu$. The data
      points are continuum subtracted on-resonance data, the
      histograms are the components of the fit, as described in the
      text.} \label{fig:2}
\end{figure*}

The amount of signal and the remaining background in the $m_{bc}$ vs.\
$\Delta E$~signal window are determined by a two-dimensional binned
likelihood fit to the lepton momentum~$p^*_l$ (4 bins, width:
300~MeV/$c$) and the invariant mass~$m(\pi^+\pi^-\pi^0)$ (8 bins,
width: 20~MeV/$c^2$)~\cite{ref:12}. The following five components are
fitted to the data: $B^+\rightarrow\omega l^+\nu$ signal,
$B\rightarrow X_ul\nu$ background, $B\rightarrow X_cl\nu$ background,
fake and non $B$~decay lepton background and continuum. The shapes of
the former four components are determined from the simulation, the
shape of the continuum component is given by the off-resonance data. The
normalizations of the $B^+\rightarrow\omega l^+\nu$, the $B\rightarrow
X_ul\nu$ and the $B\rightarrow X_cl\nu$~component are floated in the
fit, the other components being fixed (Fig.~\ref{fig:2}).

\section{Result and systematic uncertainty}

The signal yield, $N(B^+\rightarrow\omega l^+\nu)$, is determined
separately for each form factor model and the branching ratio,
${\mathcal B}(B^+\rightarrow\omega l^+\nu)$, is calculated using the
relation
\begin{eqnarray}
  \lefteqn{N(B^+\rightarrow\omega l^+\nu)=N(B^+)\times{\mathcal
      B}(B^+\rightarrow\omega l^+\nu)\times} \nonumber \\
  & & \times  {\mathcal
      B}(\omega\rightarrow\pi^+\pi^-\pi^0)\times(\epsilon_e+\epsilon_\mu)~,
\end{eqnarray}
where $N(B^+)$ is the total number of charged $B$~mesons in the
data, ${\mathcal B}(\omega\rightarrow\pi^+\pi^-\pi^0)=(89.1\pm
0.7)\%$~\cite{ref:9} and $\epsilon_e$ ($\epsilon_\mu$) is the model
dependent selection efficiency for the electron (muon) channel
(Table~\ref{tab:1}). Averaging over the three models (giving equal
weight to each), $421\pm 132$ signal events are found in the data,
corresponding to a branching fraction of $(1.4\pm 0.4(stat))\cdot
10^{-4}$. The spread between the different models amounts to 21\% and
is used as an estimate of the form factor model uncertainty.
\begin{table*}
  \begin{center}
    \begin{tabular}{|c|c|c|c|}
      \hline
      \rule[-1.3ex]{0pt}{4ex}form factor model &
      $\quad N(B^+\rightarrow\omega l^+\nu)\quad$ & $\quad {\mathcal
        B}(B^+\rightarrow\omega l^+\nu)\quad$ &
      $\qquad \chi^2\qquad$\\
      \hline
      \rule[-1.3ex]{0pt}{4ex}ISGW2 & $359\pm 106$ & $(1.02\pm
      0.30)\cdot 10^{-4}$ & 30.5\\
      \rule[-1.3ex]{0pt}{4ex}UKQCD & $428\pm 134$ & $(1.43\pm
      0.45)\cdot 10^{-4}$ & 30.6\\
      \rule[-1.3ex]{0pt}{4ex}LCSR & $476\pm 156$ & $(1.73\pm
      0.57)\cdot 10^{-4}$ & 30.6\\
      \hline
      \rule[-1.3ex]{0pt}{4ex}average & $421\pm 132$ & $(1.39\pm
      0.44)\cdot 10^{-4}$ & \\
      \hline
    \end{tabular}
  \end{center}
  \caption{The signal yield, the corresponding $B^+\rightarrow\omega
    l^+\nu$~branching fraction and the $\chi^2$ of the fit
    ($\mathrm{ndf}=31$), for each of the three form factor models used.}
    \label{tab:1}
\end{table*}

The largest experimental uncertainty is the efficiency of the neutrino
reconstruction (7.5\%). It is estimated by varying the track finding
efficiency, the neutral cluster finding efficiency, the track momentum
resolution and the cluster energy resolution separately in the Monte
Carlo within their estimated range of uncertainty. The next to largest
component is the cross-feed from decays~$B\rightarrow X_ul\nu$ (4.4\%)
which is estimated by varying the fraction of $B\rightarrow\pi l\nu$
and $B\rightarrow\rho l\nu$ (these decays are expected to dominate in
the high $p^*_l$~region) within their experimental uncertainty. Other
contributions are: $B\rightarrow X_cl\nu$~cross-feed (2.4\%), charged
track and cluster finding (3.0\% and 4.0\%, respectively), lepton
identification (3.0\%), the number of $B\bar B$~events (0.6\%) and the
uncertainty in the $\omega\rightarrow\pi^+\pi^-\pi^0$ branching
fraction (0.8\%).

\section{Conclusion}

We have measured the decay~$B^+\rightarrow\omega l^+\nu$ using
78~fb$^{-1}$ of $\Upsilon(4S)$ data (85.0~million $B\bar B$
events). The final state of the decay was fully reconstructed using
the $\omega$~decay into $\pi^+\pi^-\pi^0$ and detector hermeticity to
infer the neutrino momentum. The signal yield was determined by a
two-dimensional fit to the lepton momentum and the invariant
$\pi^+\pi^-\pi^0$~mass. Repeating the fit for three different
$B^+\rightarrow\omega l^+\nu$ form factor models and averaging the
result, $421\pm 132$ signal events are
found, corresponding to a branching fraction of $(1.4\pm 0.4(stat)\pm
0.2(syst)\pm 0.3(model))\cdot 10^{-4}$ for this decay. This result is
preliminary.

\section{Acknowledgements}

We are greatly indebted to our
funding agencies for their support in building and operating the Belle
detector, and to the KEKB accelerator group for the
excellent operation of the KEKB accelerator.

\end{document}